\documentclass[aps,pra,twocolumn,showpacs]{revtex4}
\usepackage{times}
\usepackage{amsmath}
\usepackage{amssymb}
\usepackage{graphicx}
\usepackage{epstopdf}

\newcommand{\abs}[1]{\left|{#1}\right|}

\newcommand{\expect}[1]{\langle{#1}\rangle}

\newcommand{\atan}{\mathrm{atan}}

\newcommand{\ep}{\epsilon_0}

\begin{document}

\title{Comment on ``Coupled dynamics of atoms and
  radiation--pressure--driven interferometers''} 

\author{J.\ K.\ Asb\'oth$^{1,2}$} \author{P.\ Domokos$^{2}$}

\affiliation{
  $^1$Research Institute of Solid State Physics and Optics, Hungarian
  Academy of Sciences, H-1525 Budapest P.O. Box 49, Hungary\\
$^2$Institut f\"ur Theoretische Physik,
  Universit\"at Innsbruck, Technikerstr.~25, A-6020 Innsbruck, Austria
}

\begin{abstract} 
  In two recent articles \cite{meiser06,meiser06b}, Meiser and Meystre
  describe the coupled dynamics of a dense gas of atoms and an optical
  cavity pumped by a laser field. They make two important simplifying
  assumptions: (i) {\it the gas of atoms forms a regular lattice and
    can be replaced by a fictitious mirror}, and (ii) {\it the atoms
    strive to minimize the dipole potential}. We show that the two
  assumptions are inconsistent: the configuration of atoms minimizing
  the dipole potential is not a perfect lattice. Assumption (ii) is
  erroneous, as in the strong coupling regime the dipole force does
  not arise from the dipole potential. The real steady state, where
  the dipole forces vanish, is indeed a regular lattice. Furthermore,
  the bistability predicted in \cite{meiser06,meiser06b} does not
  occur in this system.
\end{abstract}


\maketitle

In two recent articles \cite{meiser06,meiser06b}, Meiser and Meystre
describe the coupled dynamics of movable atoms, mirrors and light in a
pumped optical cavity. 
They take the effect of the atoms on the light
field into account using a 1-dimensional non-perturbative model
introduced by Deutsch et al.~\cite{deutsch95}.  
The mechanical effects of the off-resonant light on the atoms are
described by a dipole Hamiltonian. They claim that the atoms trapped
in the cavity field self-organize to form a regular lattice of atom
clouds which behaves as a beam splitter (BS). This BS effectively
splits the single cavity into two coupled resonators (``left'' and
``right'').  An important prediction of \cite{meiser06,meiser06b} is a
bistability effect: optical forces will push this ``atom BS'' to a
position where it is approximately an integer multiple of the
half-wavelength away from the left (right) cavity mirror.  Thus for a
certain parameter regime the ``left'' (``right'') cavity is on
resonance with the pump laser, and has intense light, while the other
cavity has weak field.

We argue that the model used by Meiser and Meystre
contains contradictions and errors involving the way the dynamics of
the atoms (taking place on a much shorter timescale than that of the
movable cavity mirror in \cite{meiser06}) is treated.  The model
hinges on two key assumptions. The first is a claim based on
\cite{deutsch95}: (i) {\it the gas of atoms forms a regular lattice of
  pancake-shaped clouds with lattice constant $d_0 = \lambda/2
  (1+2\,\atan \Lambda/\pi)$, and can thus be replaced by a fictitious
  ``atom BS''}.  Here $\Lambda= k \eta \alpha / (2 \ep)$ is the
dimensionless polarizability density of a cloud of surface density $\eta$
composed of atoms of polarizability $\alpha<0$, with $k=2\pi/\lambda$
denoting the free-space wavenumber of the pump laser. The second
assumption is used to find the steady state position of the atom BS:
(ii) {\it the atoms strive to minimize the total dipole potential
  $H_{\mathrm{int}}$ (eq.~(2) of \cite{meiser06}, or eq.~(3) of
  \cite{meiser06b})}. In this Comment we show that these assumptions
are inconsistent, and that (ii) has to be replaced by a formula for
the \emph{force} on the atoms. We present the required expression for
the force which allows us to determine the steady state of this
system. This does not exhibit the bistability phenomenon of
\cite{meiser06,meiser06b}.

\begin{figure}
\begin{center}
  \includegraphics[width=8cm]{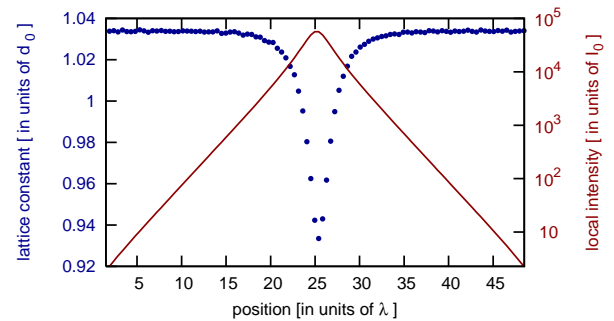}
\caption{(color online) A configuration found by Monte-Carlo
  type minimization of the dipole potential, for $N=100$ atom clouds
  with polarizability $\Lambda=0.1$, symmetrically pumped in free space
  (no cavity). The local lattice constant (blue dots, in units of the
  original $d_0=0.468\lambda$) and the electric field intensity at the
  atomic positions (red line, exponential scale, in units of the pump
  intensity $I_0$) are plotted. The total dipole energy is some 1000
  times lower than that of the regular optical lattice of assumption
  (i).  The atoms form two slabs with lattice constant exceeding
  $d_0$, impervious to the pump laser (pump frequency deep in the band
  gap, see \cite{deutsch95}).  These two self-organized atom mirrors
  constitute a high-Q cavity, trapping light in the middle.  There the
  intensity is some $10^4$ times that of the pump lasers. Inside the
  ``atom mirrors'' the intensity falls off exponentially.
\label{fig:glauber} 
}
\end{center}
\end{figure}

The setup considered in \cite{meiser06,meiser06b} is an open system:
laser light enters the cavity through the mirrors, carrying momentum
and energy, and is coupled out of the cavity at the mirrors. It is not
at all clear what function of the system parameters is minimized in a
steady state. Minimizing the dipole energy certainly does not lead to
steady state configurations. To illustrate this point, it is
worthwhile to consider a conceptually simpler situation, namely atoms
trapped in a standing wave laser field without any cavities involved.
As illustrated in Fig.~\ref{fig:glauber}, using the model of Meiser
and Meystre but allowing all of the atom clouds to move independently,
a Monte Carlo algorithm to minimize the dipole energy leads to
configurations differing from a simple lattice. Starting from the
regular lattice configuration, 
the atoms decrease the dipole energy by forming a \emph{self-organized
  cavity} resonant with the pump beam. The intensity inside this
cavity is thus enhanced by the resonance, and a few atoms coupled to
this intense field contribute to the total dipole potential by such a
large amount, that its absolute value exceeds that of the original
energy by orders of magnitude (see the figure caption for details).
The mirrors of this self-organized cavity are slabs of an atomic
lattice with lattice constant exceeding $d_0$ such that the pump field
is in the band gap \cite{deutsch95} of these slabs and decays
exponentially inside them.  The complicated spatial structure of this
solution clearly shows that assumption (ii) cannot lead to (i), one of
them has to be dropped.
The configuration represented in Fig.~\ref{fig:glauber} is not a
steady state at all: as we discuss below, radiation pressure
would push the two slabs apart.  Thus, assumption (ii) has to be
revisited.  We remark that inside an optical cavity, dipole energy
minimization leads to similar artifacts, in the ``superstrong
coupling'' limit as well.

The reason why dipole energy minimization does not supply the steady
states is that the atomic positions are coupled parametrically to the
light field. This problem is met, and is tackled in a very neat way,
when the motion of atoms coupled to a \emph{single-mode high-Q}
optical resonator is to be described in a classical approximation
(Ehrenfest theorem).  In the standard approach (e.g.,
\cite{asboth04}), the starting point is the quantum Hamiltonian
$
\hat{H} =
\sum_{j=1}^{N}\hat{H}_{\mathrm{atom}}
(\hat{p}_j,\hat{\sigma}_j,\hat{\sigma}_j^\dagger) 
+ \hat{H}_{\mathrm{field}} (\hat{a},\hat{a}^\dagger) 
+ 
\hbar\sum_{j=1}^N g(\hat{z}_j) 
(\hat{\sigma}_j^\dagger\hat{a} + 
\hat{\sigma}_j \hat{a}^\dagger),
$ with $\hat{z}_j$, $\hat{p}_j$, and $\hat{\sigma}_j$ denoting the
position, momentum, and deexcitation operator of the $j$-th atom,
$g(z)$ the mode function of the cavity, and $\hat{a}$ the cavity
photon annihilation operator.  The coupling between atoms and the
cavity field is of the celebrated Jaynes--Cummings type.  The
classical approximation should furnish equations of motion for $z_j =
\expect{\hat{z}_j}$ and $p_j = \expect{\hat{p}_j}$.  In order to
derive these, for slowly moving atoms, the separation of the
timescales is invoked.  The internal variables $\hat{a}$ and
$\hat{\sigma}_j$, which equilibrate fast on the timescale of atomic
motion, are replaced by their adiabatic steady state expectation
values, which depend on all of the atom coordinates $\hat{z}_l$ (as
well as the intensities and phases of the pumping lasers).  One could
be tempted to use the ``effective'' Hamiltonian
$\hat{H}_\mathrm{eff}(\hat{z}_j, \hat{p}_j)$ obtained from $\hat{H}$
in this way, and derive the atomic dynamics from it via the Heisenberg
equations of motion, i.e.,
$dp_j/dt=\expect{-d\hat{H}_\mathrm{eff}/d\hat{z}_j} 
=-d\expect{\hat{H}_\mathrm{eff}}/dz_j=-d\expect{\hat{H}}/dz_j$.
The correct procedure, however, is to apply the adiabatic
approximation to the original Heisenberg equations
\begin{equation}
\label{eq:heisenberg}
\frac{d}{dt}\hat{p}_j =  
\frac{1}{i\hbar}[\hat{p}_j,\hat{H}] =  
-\frac{d}{d\hat{z_j}}\hat{H}=\hat{F}_j,
\end{equation}
where the force operator is $\hat{F}_j =
-\hbar (\hat{\sigma}_j^\dagger\hat{a} + \hat{\sigma}_j \hat{a}^\dagger)
\,dg(\hat{z}_j)/d\hat{z}_j$, and use $dp_j/dt=\expect{\hat{F}_j}$.
In other words, the differentiation should only be applied
w.r.t.~\emph{explicit} $z_j$-dependence of $\expect{\hat{H}}$, since
$\expect{d\hat{H}/d\hat{z}_j}$ is not the same as
$d\expect{\hat{H}}/dz_j$.  Using $d\expect{\hat{H}}/dz_j$ to define
the dynamics supplies steady states where the ``dipole potential''
$\expect{\hat{H}}=\expect{\hat{H}_\mathrm{eff}}$ is minimized, but
these are not the true steady states of the system: in these states
the force on the atom $\expect{\hat{F}_j}$ does not vanish. 

We now turn to the setup considered by Meiser and Meystre, where the
back-action of the atoms on light is so substantial that the cavity no
longer has a fixed mode function. To be self-contained, and to fix
notation, we briefly summarize the model, detailed in
\cite{meiser06,meiser06b,deutsch95}.  

Assuming that the atoms are
fixed on the timescale of the field dynamics, the light field is
calculated by solving the Helmholtz equation in a one-dimensional
approximation,
\begin{equation}
\label{eq:helmholtz}
\partial_z^2 E(z) + k^2 E(z) = -2 k \Lambda E(z) \sum_j \delta(z-z_j).
\end{equation}
The right-hand-side embodies the polarizability of the trapped atoms,
which are assumed to form pancake-shaped clouds of axial size much
smaller than a wavelength. The solution of this equation is
trivial: between two atom clouds, the electric field is a
superposition 
\begin{multline}
\label{eq:modes}
E(z_{j-1}< z < z_j)=A_{j}e^{ik(z-z_j)}+B_{j}e^{-ik(z-z_j)}\\
=C_{j-1}e^{ik(z-z_{j-1})}+D_{j-1}e^{-ik(z-z_{j-1})}.
\end{multline}
The field has to fulfil boundary conditions:
\begin{subequations}
\label{eq:fit_E}
\begin{align}
 E(z=z_j-0) &= E(z=z_j+0);\\
\label{eq:fit_grad_E}
\partial_z E(z=z_j-0) &= \partial_z E(z=z_j+0) + 2k\Lambda E(z_j).
\end{align}
\end{subequations}
These conditions are equivalent to representing the atom clouds by
BS's, i.e., $A_j=t C_j + r B_j$, $D_j=t B_j + r C_j$ with complex
reflection and transmission coefficients $r=i\Lambda /(1-i\Lambda)$, $t=1
/(1-i\Lambda)$ \cite{deutsch95}.  

The dynamics of the atoms is given by the \emph{dipole force} acting
on them. Instead of minimizing a dipole potential, the true
\emph{steady state} of the system is then specified by the positions
of all the atom clouds $z_j, j=1,\ldots,N$ such that the optical field
of the cavity -- the solution of \eqref{eq:helmholtz} -- exerts no net
force on any of the clouds. In the following we show two ways
to calculate this force acting on an infinitely thin atom cloud (BS).

The force on an atom cloud is can be obtained by integrating the 
the force on a single atom over the whole
cloud. A microscopic model of light-matter interaction leads to two
types of force: the dispersive \emph{dipole force} and the dissipative
\emph{scattering force} \cite{tannoudji-leshouches}.  This latter is
often referred to as ``radiation pressure'', but following
Meiser and Meystre we use this term to denote the mechanical effects
of light in general.  In \cite{meiser06,meiser06b} the atom-pump
detuning is assumed to be so large that the scattering force
can be neglected, tantamount to assuming $\Lambda\in
\mathbb{R}$. For linearly polarizable particles the dipole force  
time-averaged over an optical period is
given 
\cite{tannoudji-leshouches} $F =
\frac{1}{4} \alpha \nabla\abs{E({\mathbf x})}^2$.
Calculating this force for an infinitely thin disk-shaped atom
cloud
poses a problem, as the electric field $E(z)$ is not differentiable at
the atomic positions $z_j$. One must calculate the force on a disk of
finite extent $z_j-w \ldots z_j+w$, and only then take the limit $w\to
0$.  Since the electric field is polarized in the plane of the disk,
there is no surface contribution \cite{barnett06}, and the integral in
the limit of vanishing width gives
\begin{equation}
\label{eq:force}
F_j = \frac{\eta \alpha}{8} \left(\partial_z \abs{E}^2(z_j-0)
+ \partial_z \abs{E}^2(z_j+0)\right) 
\end{equation}
for the force on a unit surface (``radiation pressure'').  This result
is independent of the way in which the limit is approached, i.e.~of
the axial density distribution of the cloud.  Substituting the modal
decomposition of Eq.~\eqref{eq:modes}, using the BS relations, and the
fact that as $\Lambda\in\mathbb{R}$, we have
$\abs{A_j}^2+\abs{D_j}^2=\abs{B_j}^2+\abs{C_j}^2$, some algebra leads
to the simple formula 
\begin{equation}
\label{eq:force2}
F_j = \frac{\ep}{2} \left(\abs{A_j}^2 + \abs{B_j}^2 -\abs{C_j}^2 
-\abs{D_j}^2 \right).
\end{equation}

There is another, macroscopic way to arrive at the light-induced force
on a scatterer.
This consists of
calculating the Maxwell stress tensor and integrating it on an
arbitrary fictitious surface enclosing the body (see, e.g.,
Ref.~\cite{antonoyiannakis}). In the 1D model of
\cite{deutsch95,meiser06}, this is very easily done.  For a selected
atom cloud, we take the surface around it to consist of two planes
orthogonal to the cavity axis, between the atom cloud and the two
neighbouring clouds. As the electromagnetic wave inside the cavity is
transverse, both the $\mathbf{E}$ and $\mathbf{B}$ vectors lie in the
planes, and the only part of the stress tensor contributing to the
integral is the term with the energy density. For the plane waves of
\eqref{eq:modes} this results in \eqref{eq:force2}. We remark that
this line of thought is also alluded to by Meiser and Meystre, and
although is not applied to the atoms, it is used to derive the force
on the movable cavity mirror in Eq.~(10) of \cite{meiser06}.

\begin{figure}
\begin{center}
  \includegraphics[width=8cm]{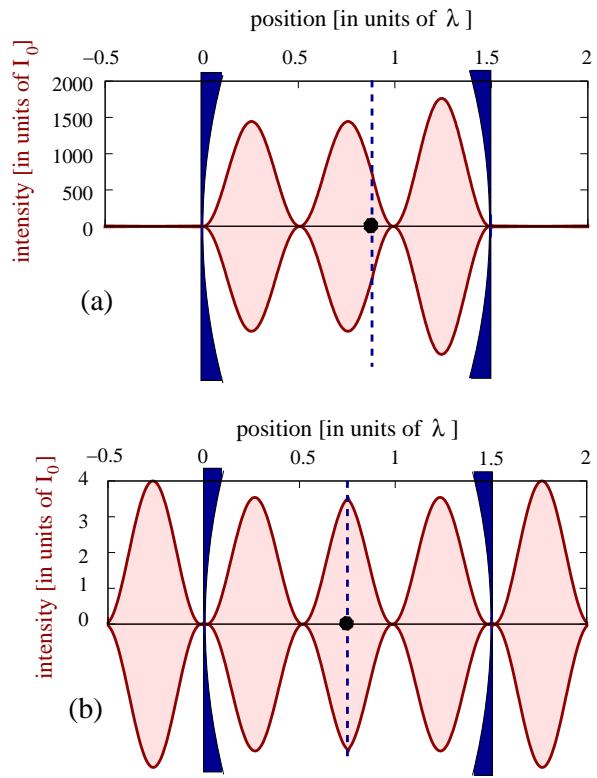}
\caption{(color online) Steady states of a single atom (atom cloud) of
polarizability $\Lambda=0.1$ in a symmetrically pumped cavity, according
to (a) minimization of the dipole energy; and (b) vanishing dipole force.
 The cavity is formed by two highly reflective mirrors
($\Lambda=10$) at $z=0$ and $z=1.501\lambda$.  The position of the atom
(black dot and vertical dotted line) induces no substantial change of the \emph{mode
function} of the cavity. However, it influences the \emph{intensity}
of the cavity mode (filled red curve, in units of the free-space
intensity $I_0$, mirrored for better visibility). In (a), the system
is on resonance, the intracavity intensity is so high (some $500
I_0$) that the intensity outside the cavity is hardly seen on this range; 
in (b), the atom is maximally coupled to the cavity, shifts it
out of resonance and reduces the intensity below the free-space value.
\label{fig:example} 
}
\end{center}
\end{figure}

The difference between (a) minimizing the dipole potential, and (b)
requiring the dipole force to vanish is illustrated in
Fig.~\ref{fig:example}.  Here we put a weakly reflective BS (
$\Lambda=0.1$) in a high-Q cavity composed of two Dirac-$\delta$
distributions of polarizability with $\Lambda=10$, corresponding to
transmission probability $T\approx 0.01$, as in
\cite{meiser06,meiser06b}.  To a good approximation the cavity only
supports a single $\sin(z)$-mode, and since the BS is weakly
reflective, we are not in the ``superstrong coupling'' limit: the
Hamiltonian approach would constitute a reasonable approximation.
Note however, that in Fig.~\ref{fig:example}(a), the intensities
to the right of the trapped atom are slightly higher than to its left,
indicating the corrections to the single-mode approximation, and also
the bistability of \cite{meiser06}.  When minimizing the dipole energy
(a), the BS occupies a position where it is not coupled strongly to
the cavity, far from an antinode of the mode function.  Thus the
system is near-resonant with the driving field, and the intracavity
intensity is enhanced by the resonance, some 500 times the intensity
outside.  Extremely high intensity means large negative dipole energy
for the trapped atom -- however, note that at the position of the
atom, the derivative of the intensity is nonvanishing, and thus the
force $F$ is nonzero.  Requiring this \emph{force} to vanish brings
the atom to a mode function antinode (b), where the coupling is
stronger, hence the frequency shift is larger, than in (a). For this
specific example, this is already enough to shift the system out of
resonance and to decrease the intracavity intensity below the
free-space value.

\begin{figure}
\begin{center}
  \includegraphics{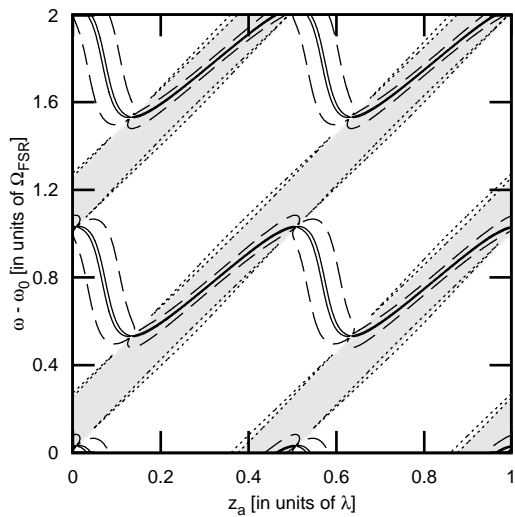}
\caption{Force on a single beam splitter with $\Lambda=1$ in  
  a standing-wave cavity with parameters as in
  \cite{meiser06,meiser06b} (mirror transmission probability $T\approx
  0$) pumped by a laser via one of the end mirrors. Contour lines are:
  solid, $\abs{F}=10 F_0$; dashed, $\abs{F}= F_0 / 10$; dotted,
  $\abs{F}=F_0/1000$, where $F_0$ is the radiation pressure force that
  would act on the beam splitter in the absence of the cavity. Gray
  background indicates positive (rightward) forces, white background
  negative (leftward) forces.  Extremely high forces ($\abs{F}>10F_0$,
  the small areas enclosed by the solid contour lines) occur whenever
  a part of the cavity is on resonance with the drive. The boundaries
  between the gray and white areas, where $F\approx 0$, are the
  (stable or unstable) equilibrium positions.
\label{fig:map} 
}
\end{center}
\end{figure}

The BS used to model the atom cloud in the cavity considered by
\cite{meiser06,meiser06b}, has an effective $\Lambda\approx 1$, and
thus the system is in the ``superstrong coupling limit''.  We plot the
force exerted by the cavity field on such a BS as a function of its
position and of the detuning (equivalently, cavity size) 
in Fig.~\ref{fig:map}. The boundaries between the gray shaded and
white areas correspond to equilibrium.  Note however, that for a fixed
drive detuning, of the two equlibrium solutions per half wavelength
only one is stable, the one where $\partial/\partial z_a \, F <0$.
Thus there is no bistability of the kind predicted in \cite{meiser06}.
The areas enclosed by the solid contour lines, where the force on the
BS becomes very large, indicate that the system is on resonance and
the field gradient at the BS position is high. These correspond to the
black structures on Fig.~2.~of \cite{meiser06b}, where the determinant
$D$ (Eq.~(12) of \cite{meiser06b}) becomes small, which there are
falsely interpreted as equilibrium positions.

In the example shown in Fig.~\ref{fig:example}, the maxima of the
intensity to the left and to the right of the trapped BS are equal. In
fact, an analogous statement holds for the steady state of any
one-dimensional system composed of $N$ consecutive beam splitters,
held together by the dipole force, regardless of the BS parameters of
the system components. This follows from formula~\eqref{eq:force2} for
the dipole force, whereby for every $j=1,\ldots,N$:
$\abs{A_j}^2+\abs{B_j}^2=\abs{C_j}^2+\abs{D_j}^2$.  Furthermore, since
there is no absorption,
$\abs{A_j}^2+\abs{D_j}^2=\abs{B_j}^2+\abs{C_j}^2$ at every BS.  These
relations imply that $\abs{A_1}=\abs{A_2}=,\ldots,=\abs{A_N}=
\abs{C_1}=\abs{C_2}=,\ldots,=\abs{C_N}$;
$\abs{B_1}=\abs{B_2}=,\ldots,=\abs{B_N}=\abs{D_1}=\abs{D_2}
=,\ldots,=\abs{D_N}$, i.e., the plane waves pass the atom clouds
unattenuated, suffering only phase shifts.  Thus, the envelope of the
intensity oscillations of the electric field is constant throughout
the sample. This is in line with the intuitive picture of ``radiation
pressure'' caused by the collisions of photons with the atom clouds.
However, these results are in direct contradiction to those obtained by
Meiser and Meystre: the nonconstant field envelopes in the
``bistability regime'' are explicitly plotted in Fig.~3 of
\cite{meiser06b}.

We now revisit assumption (i).
Assuming that there exists a steady state of $N$ identical disk-shaped
atom clouds trapped by the light field in a cavity, we find that 
these clouds have to form a perfect lattice.
This is true because in the steady
state
 the light permeates the
stack of clouds unattenuated. Thus
$\abs{E(x)}^2=\abs{E_0}^2+\abs{E_1}^2 + 2 \abs{E_0
  E_1}\cos(2kx-\Phi(x)) $ everywhere in the sample, the clouds only
contribute to the phase: $\Phi(x_j<x<x_{j+1}) =\sum_{l=1}^{j} \chi_l
$. The phase slip at the $l$'th cloud $\chi_l$ depends not only on the
polarizability density $\Lambda$ of the clouds, but also on the pump
asymmetry, i.e., the ratio of the intensities of the left- and
rightwards propagating waves, see \cite{asboth06} for details. Since
the clouds are identical, and the light fills the structure
unattenuated, both these parameters are equal for all clouds.
Thus, $\chi_l=\chi$ for every $l=1,\ldots,N$, therefore the atom
clouds form a perfect lattice (possibly with gaps of an integer
multiple of $\lambda/2$) with lattice constant $d =
\frac{\lambda}{2\pi}(\pi-\chi)$.  However, the value of the phase slip
$\chi$, and thus of the lattice constant $d$, is not trivial to
determine, due to the dependence on the pump asymmetry. For the atoms
trapped inside the asymmetrically pumped cavity considered by Meiser
and Meystre, this asymmetry varies with the pump detuning: on
resonance, it is negligible, while far from resonance, it is
substantial.  Thus Eqs.~(24) and (25) of \cite{meiser06} (taken from
\cite{deutsch95} for symmetric pumping) \emph{cannot be applied} to
this system.
Even more crucially, whether or not the equilibrium
 sets in depends on
the \emph{dynamics}.  We have found \cite{asboth06} that for large
lattices, even a small pump asymmetry can lead to a dynamical
instability of the equilibrium configuration.

Common wisdom holds that ``dipole force is conservative''.
Even in the strong coupling regime of cavity QED it is often possible
to construct a ``potential'' by integrating the dipole force (e.g.,
\cite{asboth04}).  For many particles trapped in the same cavity, this
potential should also include the field-mediated (parametric)
interaction between those particles.  We have found \cite{asboth06},
that in the generic case even this approach breaks down. For
asymmetric pumping (as in \cite{meiser06,meiser06b}), $\partial
F_j(z_1,\ldots,z_N)/\partial z_l \neq \partial
F_l(z_1,\ldots,z_N)/\partial z_j $, and thus no potential function
$V(z_1,\ldots,z_N)$ can be constructed that obeys the Young theorem
about the commutativity of partial derivatives.

This work was supported by the National Scientific Fund of Hungary
(Contract Nos.~T043079, T049234, NF68736) and the Austrian Science
Foundation (P17709 and S1512).

\end{document}